\newcommand{\beq}{\begin{equation}}
\newcommand{\eeq}{\end{equation}}

\documentclass[aps,preprint,pra,showpacs]{revtex4-2}
\usepackage{graphicx}
\usepackage{dcolumn}
\usepackage{bm}
\usepackage{amsmath}
\usepackage[colorlinks =true,
            linkcolor = blue,
            urlcolor  = blue,
            citecolor = blue,
            anchorcolor = blue]{hyperref}
\usepackage{amssymb}
    \usepackage{braket}
\usepackage{subcaption}
\usepackage{multirow}

\begin{document}

\title{Quantum Test of the Local Position Invariance with Internal Clock Interferometry}%
\author{Zhifan Zhou}%
\email[Email: ]{zhifan@umd.edu}
\affiliation{Joint Quantum Institute, National Institute of Standards and Technology and the University of Maryland, College Park, Maryland 20742 USA}
\date{1/26/2023}%

\begin{abstract}
%

Current attempts to test local position invariance\, (LPI) compare different clock 
transition rates with classically exchanged signals. We propose an experimental scheme for the quantum test of LPI: an internal atomic clock interferometer comprising two interfering clocks within one atom. We prepare the atom in a superposition of two clock states and one ground state, which evolves coherently along two quantum clock oscillations into stable internal Ramsey interference patterns. The interference pattern with the shared ground state shows a visibility modulation, which can be interpreted as the beating of the individual clock oscillations and a direct consequence of complementarity. Upon the interferometer experiencing a different gravitational potential, LPI predicts that both clock tick rates will change proportionally, while quantum complementarity indicates that 
the visibility modulation should modify accordingly. This change is deemed insignificant for the first period of visibility modulation but can be stacked up until the limit of the system coherence time. Since no splitting or recombining is involved, the system coherence time can be as long as the trap lifetime or the clock state lifetime. The required resolution to observe the visibility modulation is within reach of the state-of-art optical clocks' sensitivities. This experimental scheme is feasible in different scenarios, still or with speed, and may shed new light on studying the quantum effect of time and general relativity. 

\end{abstract}
\maketitle

Local position invariance (LPI), together with the weak equivalence principle (WEP) and local Lorentz invariance (LLI), services as a prominent part of the Einstein equivalence principle (EEP). A standard test of LPI typically involves comparing different clock transition rates at the same location~\cite{LPI2007a, LPI2007b, LPI2008, LPI2012, RMP15} or the changed transition rates with the same clock at different locations~\cite{Wineland10, Galileo18a, Galileo18b, Katori20, EEP2021}. Recent years have witnessed more involved clock transitions, including the two transitions within the same atom~\cite{Safronova18, Peik21}. Yet the development of quantum interferometry has brought new insight into the interrogation of EEP in the regime in which quantum mechanics (QM) becomes relevant~\cite{Zych11, Pikovski2015, ClockScience, Zych16, NJP2017, QuCom}. Motivated by the subtle role that the quantum effect may play, we have sent two different clock transitions that coherently interfere with each other within a single atom. Our proposed internal clock interferometry can serve as a new tool for studying the interplay of general relativity (GR) and QM.

In the standard test, LPI is done by comparing the rates of classic clocks within the gravitational fields by classical exchange of electromagnetic signals. Hence, no quantum interferometric visibility is observed, resulting from the which-path witness and complementarity. In the quantum test of LPI, the two clock branches accumulate different phases with different evolving rates. From the view of quantum mechanics, paths become distinguishable, resulting in visibility modulation. Consequently, whereas standard clock comparisons test the classical LPI, internal clock interferometry probes the quantum aspects of LPI. For example, under different gravitational potentials, the absolute difference of the clock transition frequencies will be changed, and whether the visibility modulation changes with the fundamental frequency difference will tell whether quantum complementarity holds in this gravitational redshift regime. On the other hand, standard two-slit interference stands at the heart of standard quantum mechanics, where the paths are not sensitive to redshift and gravitational fields. Internal clock interferometry can be treated as a varied version of two-slit interference with paths coupled with gravitational fields. The more involved clock states will lead the platform for testing three-slit interference under the influence of gravitational effects\,\cite{Science10, Samuel14, Leuchs16, Dima2022}. 

It is worthwhile noting the decades of progress in precise atomic clocks\,\cite{Wineland10, RMP15, Ye2022, Shimon2022}, optical clocks in space~\cite{KBongs, Space2020, Marianna2022}, and matter-wave interferometry\,\cite{Tino2007, Hogan2013, Mark15, MAGIS2021}. Specifically, we note the recent proposals towards observing visibility modulation with the spatial superposition of clock wave packets\,\cite{Zych11, Pikovski2015, NJP2017, Giese21}. The proposal has been realized with a prototype experiment with an atom chip system\,\cite{ClockScience, QuCom, GeoPhaseJump} involving an emulated time lag. 

In our experiment, internal atomic clock interferometry - atoms in superpositions of two clock states - involve an internal Ramsey scheme (Fig.1). We investigate that the visibility of population oscillation depends on the relative clock rotation, the proper time differential between each branch of the clock evolution. Our proposal shows that a differential clock reading affects the visibility of internal clock interferometry; specifically, the visibility equals the scalar product of the interfering clock states. 

In principle, any system evolving with two well-defined periods can be used as internal clock interferometry. Here we used a quantum three-level system. The candidates include neutral atoms, ions, and nuclear clocks. In our proposed experiment, the internal clock interferometer - clocks that interfere coherently within one atom - are prepared as a coherent superposition of a three-level system. To maximize the interferometric effect, the population ratios among the two excited clock states and the ground state are prepared as 0.25, 0.25, and 0.5. As the clock transitions at each branch differ, atoms accumulate phases at different rates. Thus each branch measures proper time and contains which-way information, leading to visibility modulation. 

To examine the coherent interference of internal clock superposition, we let the two clock transitions be freely involved and closed by another $\pi$/2 pulse to create internal fringes. 
Because the phase of $\pi$/2 is tunable to reveal the Ramsey fringe, the phase can be scanned to test the system's coherence. We note that during the interrogation process, the atom is trapped and experiences no splitting or recombining processes. Hence, the system coherence time is not influenced by the splitting and recombining atomic wave packets in standard atom interferometers. The upper limit will be the clock state lifetime or the trap lifetime.

We now show that visibility modulation is observed as a function of interrogation time. We treat the $|1 \rangle$ to $|3 \rangle$ transition as clock 2 and the $|1 \rangle$ to $|2 \rangle$ transition as clock 1. After the initialization, the atom is equally split into two branches. Clock 2 has a larger transition frequency than clock 1, and this branch accumulates phase faster than the other. The phase difference will be reflected in the output as the population of state $|1 \rangle$. In the extreme case, the two clocks are orthogonal to each other - for example, one is in the state of $|1 \rangle$-$|3 \rangle$, and the other is in the state of $|1 \rangle$+$|2 \rangle$. The visibility of the state $|1 \rangle$ population will drop to 0. This is contrary to a standard matter-wave interferometer, where the visibility is always high. A revival of the visibility is seen when the differential rotation angle is involved further than $\pi$ and reaches 2$\pi$. The periodicity of this visibility modulation will be 1/$\Delta f$. The beat frequency out of the clock interference is $\Delta f = f_2 - f_1$. The population of state $|1 \rangle$ oscillates with the function as:
 \beq P = \frac{1}{2}[1+cos(\Delta f t/2)cos(\omega t)], \eeq 
 The visibility reads:
 \beq V = |cos(\Delta f t/2)|. \eeq


To obtain a more general view of the effect, we studied the dependence of the interference pattern visibility on different gravitational potentials by putting the interferometer further away. For example, one is located on the earth's surface, and the other one's position is changed by $\Delta h$ (Fig.2). The clock transition frequencies and the beat frequency are shifted by the fractions of $ \delta f_1/f_1 =g\Delta h/c^2$, $\delta f_2/f_2 = g\Delta h/c^2$, and $\delta (f_2 - f_1)/(f_2-f_1) = g\Delta h/c^2$.

The essence of the which-path witness is that visibility is complementary to the clock's relative rotation. In Fig.2, we chose to work with the first oscillation period to highlight the visibility's dependence on the proper clock rotation's change. The time change for reaching the minimal visibility for the first period is insignificant as $1/\Delta f \times g\Delta h/c^2$. To further increase this signal to make the change from the gravitational field resolvable, we can extend the interrogation time until the limit of the clock state lifetime or the trap lifetime, as shown in Fig.3. The enhancement factor is $\tau \times \Delta f$. Then the total time shift signal at the limit of system coherence time is $1/\Delta f \times g\Delta h/c^2 \times\tau \times \Delta f = \tau \times g\Delta h/c^2 $. Here we use the system coherence time 1s as one example. The visibility modulation shift for the lifting height of 1m will be $1.1\times10^{-16}$, well within reach of the state-of-art clock technique. 




The challenge of realizing such an internal clock interferometry is having two clock states with long clock lifetime as shown in \cite{Safronova18,Peik21}. The other challenge for neutral atoms or molecules is the trapping for both clock states at magic wavelengths \cite{Safronova18}. The charged ion and nuclear clocks provide appropriate candidates for providing parallel clock transitions with long clock lifetime \cite{RMP2018, Safranova2014, PRA2014, Peik2018, peik2021}. 

In conclusion, this internal clock interferometry scheme uses an interferometer to measure the relative frequency difference. Thus, it can suppress systematic common noise. Meanwhile, it doesn't suffer from the recombining issue like the usual atom interferometers do, so the setup can be located anywhere with any speed, considering the significant progress with the compact atomic clock in space~\cite{KBongs, Space2020, Marianna2022}. This internal atom interferometer can serve as a precise quantum sensor for tests of local position invariance and measurements of the possible time variation of fundamental constants such as fine-structure constant \cite{Safronova18,Peik21}. 

\begin{figure}[htp]
\begin{center}
\includegraphics[width=16.5cm, height=9.0cm]{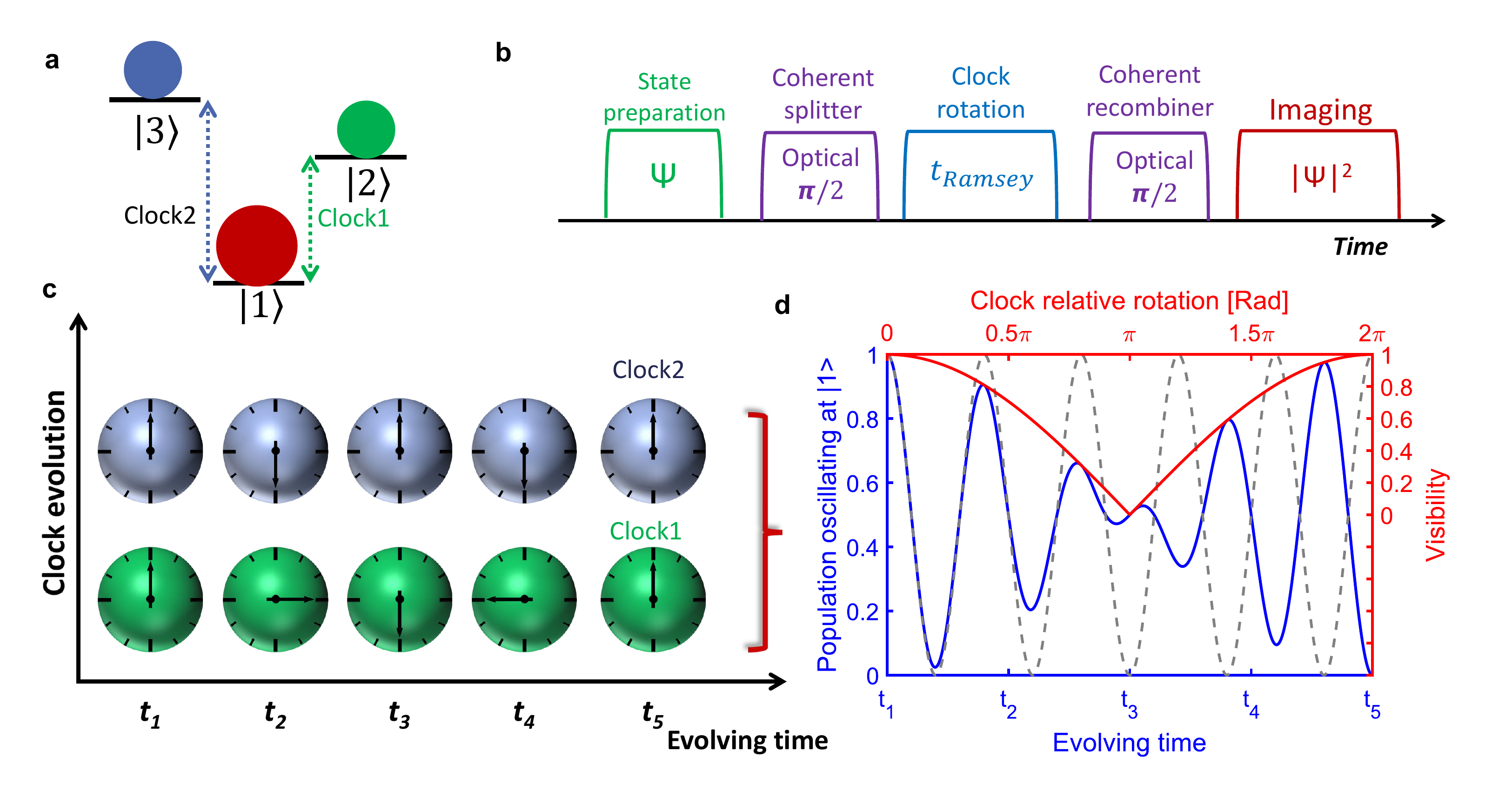}
\end{center}
\caption{ Experimental sequence of the internal clock interferometer. (a) The system consists of two clock states and one ground state. (b) Detailed sequence (not to scale). The clock interferometer is initialized by a pair of optical $\pi/2$ pulses, after which the two clocks are ticking at different rates, and the relative rotation is accumulated. The other pair of optical $\pi/2$ pulses are applied to close the interferometer. Last, the population distribution of these three states is measured. (c) Evolution in time. Each clock wave packet shows a one-handed clock, in which the hand
corresponds to a vector in the equatorial plane of the Bloch sphere. When the clock readings (the position of the clock hand) in the two clock wave packets are the same, the fringe visibility is high. When the clock readings are opposite (orthogonal), there is a "which path" witness, and the fringe visibility is low. (d) the visibility evolution in time, synchronized with (c). }
\label{fig-Scheme}
\end{figure}

\begin{figure}[htp]
\begin{center}
 \includegraphics[width=18.0cm, height=13.8cm]{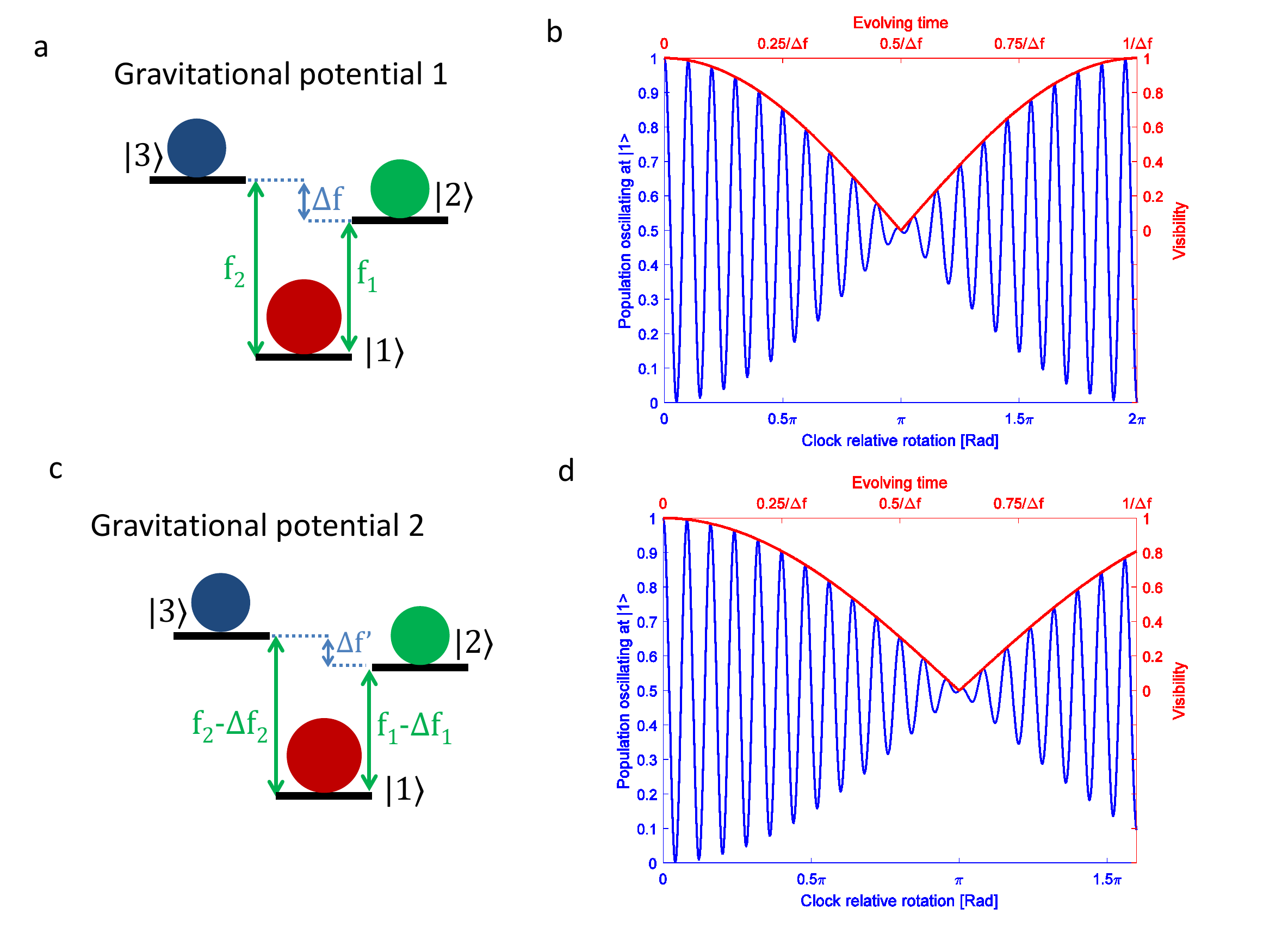}
\end{center}
\caption{ (a) When the experiment is done at the earth's surface, the beat frequency is the difference $\Delta f=f_2 - f_1$ between the two clock transitions. (b) The population of the shared state shows a visibility oscillation. (c) When the experiment is done in a different gravitational potential, for example, changed by one meter. The time dilation is $\frac {\Delta f}{f}=\frac {g \Delta h}{c^2} =\frac {9.8\times1}{9\times10^{16}}=1.1\times10^{-16}$\,(to lowest order in $c^{-2}$).
 (d) The visibility oscillation frequency will experience a shift due to the different gravitation potential. The shifted amounts are for illustration purposes, with $\Delta f$ changed by 0.2.}
\label{fig-Scheme}
\end{figure}

\begin{figure}[htp]
\begin{center}
 \includegraphics[width=14.5cm, height=10.8cm]{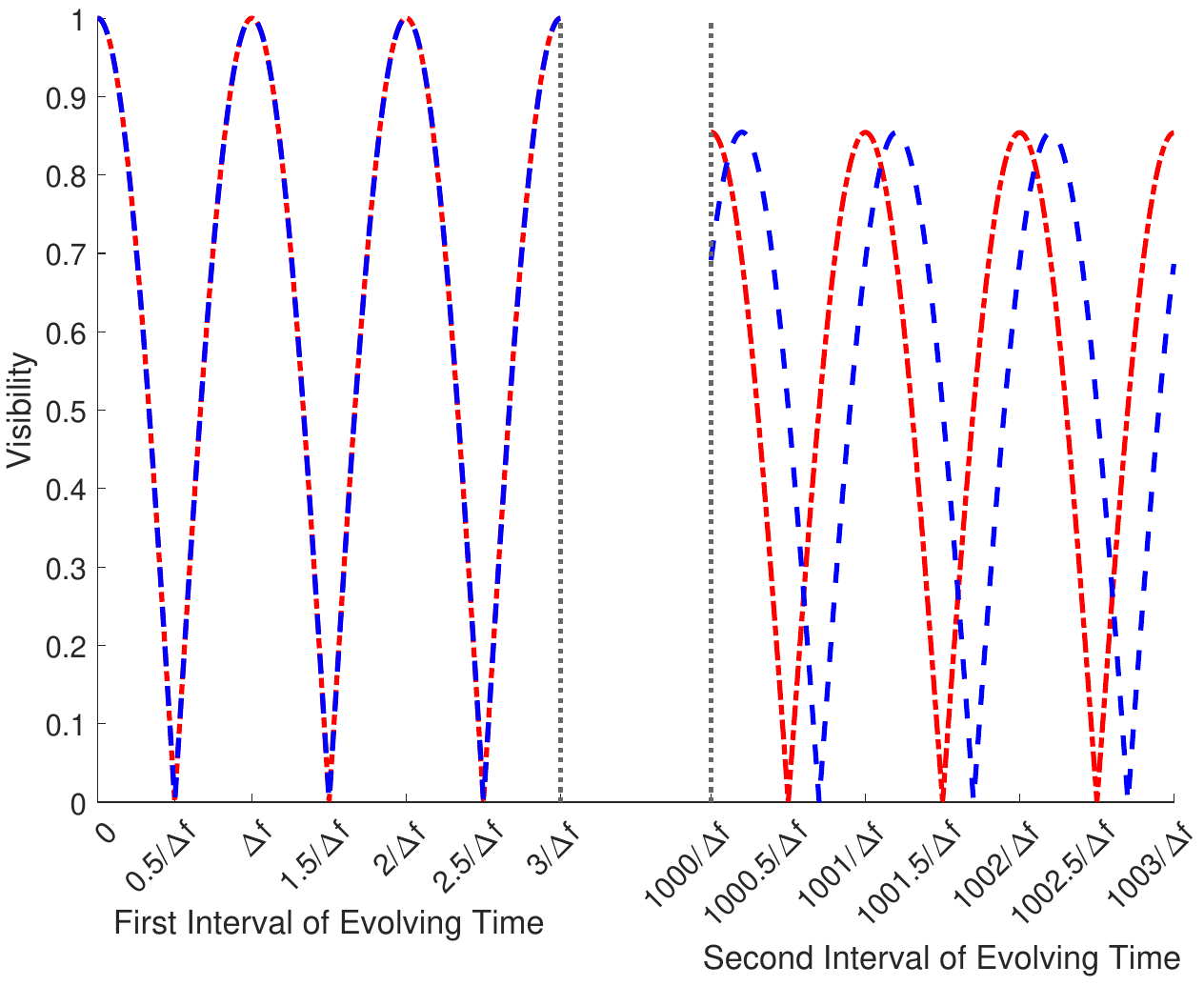}
\end{center}
\caption{The entire evolving time of the internal clock interferometry can be as long as the trap lifetime, which lasts tens of seconds. The detect resolution is not limited by the first dip of the interference visibility. Instead, the signal can be stacked up to multiple accumulating regimes. In the figure, the red and blue dash lines represent the fractional shift of $4\times10^{-4}$, which is negligible for the first several periods. After stacking for 1,000 periods, the fractional shift is 0.4. The number is not scaled with the real fractional shift.}
\label{fig-Scheme}
\end{figure}


This work was supported by the Air Force Office of Scientific Research (FA9550- 16-1-0423). We acknowledge Marianna S. Safronova for the fruitful discussions. 

\bibliography{sample}

\end{document}